
\def\av#1{\left<#1\right>}
\def\qav#1{\overline{#1}}

\title{Geographic speciation in the Derrida-Higgs
model of species formation}

\author{F Manzo\dag\ and L Peliti\ddag\footnote\S{\rm 
Boursier Henri de Rothschild. Present
address: Dipartimento di Scienze Fisiche and Unit\`a INFM,
Mostra d'Oltremare, Pad.~19, I-80125 Napoli (Italy). Associato
INFN, Sezione di Napoli.}}

\address{\dag\ Dipartimento di Fisica, Universit\`a ``La Sapienza'',
Piazzale Aldo Moro~2, I-00185 Roma (Italy)}

\address{\ddag\ Institut Curie, Section de Physique et Chimie,
Laboratoire Curie,\break 11,~rue~P~et~M~Curie, F-75231 Paris Cedex 05
(France)}

\shorttitle{Geographic speciation}


\jnl{\JPA}

\date

\beginabstract
We consider the Derrida-Higgs (DH) statistical
model of species formation in the case where the population
is geographically distributed in discrete locations, and mating
only takes place within one location.
Keeping the rate of migration
between neighbouring locations at a fixed
value, we change the mutation rate, changing
therefore the average overlap between genotypes.
When the overlap between individuals
living in different locations falls below a fecundity threshold,
speciation occurs.
When more species coexist, the
genetic structure of the population (as described by the
overlap distribution $P(q)$) fluctuates. However, the average overlap,
both within one location and among neighbouring locations,
appears to vary according to the same laws as in the absence
of speciation. The model provides a reasonable estimate 
of the parameter values necessary to observe geographic 
speciation, which is found to be much more likely than the
sympatric speciation of the original DH model.
Applications to the case of circular invasion, where
the concept of biological species appears to run
into difficulties, are sketched.
\endabstract

A number of statistical models of evolving populations
in neutral landscapes have recently been discussed in the literature
(Derrida and Peliti 1991, Serva and Peliti 1991, Derrida
and Higgs 1991, 1992). The interest of this approach lies
in the possibility to clarify evolutionary phenomena
by means of methods and concepts developed in statistical mechanics.
In this context the model introduced by Derrida and Higgs (1991,
hereafter referred to as DH) is particularly interesting,
since it may be considered as a minimal model for the
formation of species and as a kind of laboratory in which
mathematical problems related to the concept of biological
species can be formulated.

Although the model can reproduce some features of species dynamics,
many simplifying hypotheses introduced in DH appear too drastic.
The aim of the present paper is to discuss some properties
of a generalisation of DH, in which the geographic distribution
of the evolving population is taken into account. We find that,
in agreement with the present understanding of the mechanisms of
speciation, even a moderate amount of geographic 
isolation leads to species formation: it follows that the mechanism is
much more likely to act than the sympatric speciation
mechanism of the original DH model.

The main result of DH is that big populations with high mutation
rates cannot stably remain support a single species situation.
This kind of situation does not appear to occur in nature and
crucially depends on the selective mechanisms which
prevent two individuals from successfully mating.
It is usually held that, in sympatric speciation, these
mechanisms depend on few genes which would then obviously
be submitted to strong selective pressure. We think therefore
that the mechanism considered in DH, which puts all genes
on the same footing, cannot accurately describe the speciation
process.
On the other hand, the situation of coexisting species is well
documented in nature, and geographic speciation is acknowledged
as the main mechanism leading to species formation. 

In our model, as well as in DH, the number of effectives of each
species (and indeed the number of species) fluctuates widely,
and species are formed or become extinct continuously, but
the time scales involved are different. The time scales involved
in our geographic speciation model are more easily put
in correspondence with palaeontological data.

On the other hand, the average properties of the distribution of the
population in genetic space appear to vary smoothly as the driving
parameters (essentially the mutation rate) are changed, and the
system goes from a one-species to a many species regime.

For completeness, we start by a short review of the DH model.
One considers an evolving population whose number of
individuals is fixed to be equal to $M$. The genotype of
each individual is identified by the state of $N$ binary
variables $S_i^{\alpha}=\pm 1;\ \ i=1,2,\ldots,N;\ \alpha=1,2,
\ldots,M.$ The population evolves according to a reproduction-mutation
mechanism with recombination, defined by the following protocol:
\smallskip
\item{(i)} at each time step, for each individual $\alpha$ of
the new generation, one chooses two parents $\alpha^{\prime},\;
\alpha^{\prime\prime}$; the choice is random, but
constrained by the genetical closeness requirements that will be
discussed later;
\item{(ii)} for each unit $i$ of the new individual $\alpha$,
the state of the corresponding variable $S_i^{\alpha}$ is given
by the stochastic equation
$$
S_i^{\alpha}(t+1)=\varepsilon_i^{\alpha}(t)\left[\xi_i^{\alpha(t)}
S_i^{\alpha^{\prime}}(t)+\left(1-\xi_i^{\alpha}(t)\right)
S_i^{\alpha^{\prime\prime}}\right].\eqno(1)$$
Here the random variable $\xi_i^{\alpha}(t)$ takes on the values
0 and 1, each with probability ${{1}\over{2}}$, and represents the
effects of recombination, whereas mutations are represented
by the variable $\varepsilon_i^{\alpha}(t)=\pm 1$, which
satisfies $\overline{\varepsilon_i^{\alpha}(t)}=\e^{-2\mu}$,
where $\mu$ is the bare mutation rate. All these 
random variables are extracted independently at each generation,
for each individual and each genome unit.
\smallskip
Our interest lies in the evolution of the genetic structure 
of the population, which may be described by the distribution 
of the genetic overlaps
$$
q^{\alpha\beta}=\frac{1}{N}\sum_{i=1}^N S_i^{\alpha}S_i^{\beta},
\eqno(2)$$
for any pair of individuals $(\alpha,\beta)$. One can directly take
the $N\to \infty$ limit, and consider the evolution
of the overlap matrix $q^{\alpha\beta}$ without explicitely
describing the genotypes of the individuals. In this
way the sampling fluctuations introduced by the mutation mechanism
vanish.

The main feature of DH is the constraint of genetic closeness
on a pair to be able to mate successfully:
one requires that $q^{\alpha^{\prime}
\alpha^{\prime\prime}}$ be larger than a threshold $q_0$.
In practice, one chooses at random the first partner, and then the
second one is chosen among the individuals $\beta$ which satisfy 
$q^{\alpha^{\prime}\beta}\ge q_0$.

In the absence of this constraint, the population reaches a
regime in which the overlap between any two individuals
is strongly peaked around the average
$$
Q^*(M)=\frac{1}{1+M\left(\e^{4\mu}-1\right)},\eqno(3)$$
in the large population limit (Serva and Peliti 1991).
If $Q^*(M)>q_0$, the constraint is ineffective; on the other hand,
if $Q^*(M)<q_0$, after a short transition time, the population
breaks into several subpopulations (species). The overlap
between individuals belonging to the same species is
larger than $q_0$, whereas that between individuals
belonging to different species is smaller than
$q_0$. The size of each species fluctuates from
generation to generation; if at any given generation
$t$ a given species has size equal to $m$, the size of 
the species at the next generation is binomially
distributed, with a probability $(m{/}M)$.
The size of a species also determines the mutual
overlap among its members, which is given by $Q^*(m)$,
up to fluctuations which become smaller and smaller
for larger species size.
The minimum number of coexisting species is given
by $\nu=M/m^*$, where $m^*$ satisfies the equation
$
Q^*(m^*)=q_0.$

In our generalization of the DH model, the population is
distributed on geographical units, which we call islands.
Each island sustains a population of $M$ individuals.
The geographic closeness constraint requires that
the two parents of each individuals of the new generation
in each island must belong to the same island. However,
before each reproduction step, each individual can migrate
from each island to a neighboring one with a small
probability $\epsilon$. Therefore the population
$M$ is allowed to fluctuate before the reproduction step,
but comes back to $M$ after it. However, these fluctuations are
negligible for large populations.

We now consider the case of two neighbouring islands.
If one neglects at first the genetic closeness constraint, 
it is easy to derive the distribution of the genetic overlaps.
Let us denote by $Q$ the average overlap between
individuals belonging to the same island, and by $P$ the one relative
to individuals belonging to different islands.
These quantities satisfy the equations
$$\eqalign{
Q&=\e^{-4\mu}\left[\frac{1}{M}+\left(a(\epsilon)-\frac{1}{M}\right)
Q+b(\epsilon)P\right],\cr
P&=\e^{-4\mu}\left[b(\epsilon)Q+a(\epsilon)P\right],}\eqno(4)$$
where
$$\eqalign{
a(\epsilon)&=(1-\epsilon)^2+\epsilon^2;\cr
b(\epsilon)&=2\epsilon(1-\epsilon)=1-a(\epsilon).}\eqno(5)$$
We have therefore
$$\eqalign{
Q&=\left(\e^{4\mu}-a(\epsilon)\right)/\Phi,\cr
P&=b(\epsilon)/\Phi,}\eqno(6)$$
where
$$\Phi=M\left[\left(\e^{4\mu}-a(\epsilon)\right)^2-b^2(\epsilon)\right]
+\e^{4\mu}-a(\epsilon).\eqno(7)$$
The ``infinite population limit'' (Derrida and Peliti 1991) is defined
by $M\to\infty$; $4\mu M\to \nu$; $\epsilon M\to\sigma$. In
this limit we obtain
$$\eqalign{
Q&=\frac{\nu+2\sigma}{(2\nu+1)\sigma+2\nu};\cr
P&=\frac{2\sigma}{(2\nu+1)\sigma+2\nu}.}\eqno(8)$$

As long as both $P$ and $Q$ are above threshold, these results
hold and fluctuations are negligible. This may be obtained by 
an explicit calculation of the fluctuations, following, e.g.,
Serva and Peliti (1991), or by the
following argument. Assume that fluctuations are small, and that
the average overlaps are above threshold.
Then the large majority of pairs in  each island are fecund.
Take two individuals, $\alpha$ and $\beta$, whose overlaps with
a given genotype are given by $q^{\alpha}$ and $q^{\beta}$
respectively. Then (neglecting the
effect of mutations) the overlap of the offsprings of our pair 
with the given genotype will be given by the average of
$q^{\alpha}$ and $q^{\beta}$. This is a case of blending inheritance:
it is no surprise to find it here, since the overlap depends on a great
number of independent traits. In this way the width of the
distribution of the overlaps is halved at each generation,
and a steady state
is eventually reached where the average overlap is determined by
the balance between the centripetal effect of blending and the
centrifugal effect of mutations.

The situation is different if $P<q_0<Q$. In this case he two islands 
will tend to host populations which are not mutually fecund:
two different species. In principle, this could lead to a situation in which
the overlap within one island remains above threshold, with
a finite value $Q$, and the mutual overlap between the
islands eventually vanishes. In practice this does not
happen, since individuals originating from one island colonize the other
from time to time. It turns out from our simulations that
the average overlap within one island and between the two
islands follow more or less the equations written above:
however, the overlaps exhibit fluctuations, which first appear
when $P=q_0$ and increase as $P$ decrease.

This is shown in fig.~1. The average overlaps $Q$ and $P$ are
plotted against the mutation rate $\mu$ for a population of
$M=200$ individuals per island. The migration rate $\epsilon$
is set to 0.05. The threshold $q_0$ is set to 0.2. In the same plot
the expected values in the absence of the threshold are
reported, on the basis of eqs.~(6,7). The averages
are computed over $10\,000$ generations.

As stressed by Derrida and Higgs (1991) one may consider
two kinds of fluctuations: the fluctuations of the
overlap within one population are measured by
$$
\delta=\qav{\av{q^2}-{\av{q}}^2}.\eqno(9)$$
Here as in Derrida and Peliti (1991) the angular brackets denote
population average at a given generation,
while the bar denotes the average over successive generations.
On the other hand
$$
\Delta=\qav{{\av{q}}^2}-{\qav{\av{q}}}^2\eqno(10)$$
measures the fluctuations of the average overlap from generation
to generation. In fig.~2 we report these quantities as a function
of $\mu$ for the same situation as in fig.~1. Both quantities
vanish, up to finite-population corrections,
as long as $P$ remains larger than $q_0$, while they increase rapidly
as soon as $P$ falls below threshold. One can remark that $\delta$
is consistently smaller than $\Delta$, indicating that for
most of the time a well-defined species occupies each island,
and it is refreshed by immigration from the other island.
It is interesting to remark that no new phenomena appear
when $Q$ falls below threshold, since a regime with coexisting
species has already set in.

The geographic speciation mechanism is much more effective
than the sympatric one. If we set $q_0\sim 10^{-2}$ and
assume $\epsilon\sim 1/m$, then with a reasonable mutation
rate $\mu\sim 10^{-8}$ the minimum number $M$ of individuals
to observe geographic speciation is of order $10^3$, against
the $10^6$ necessary for sympatric speciation. Let us
remark that for such a large population the hypothesis
of panmicticity (i.e., that any two individuals
have the same probability of mating) is hardly acceptable. 

We plot in fig.~3 the number $\nu$ of potentially fecund
partners per individual as a function of time in a typical
run for this system ($M=200;$ $\epsilon=0.05;$
$\mu=0.0025$). Initially all individuals are identical
and this number equals 400. After a certain time
speciation occurs as it may be deduced by the fact that
this number can take on two values, which are both close to
the number of inhabitants of one island.
However, from time to time one observes more complicated
speciation events, which are represented in this
plot by the coexistence of several different
values of $\nu$. Around generation 800 one
observes the coexistence of {\it three\/}
species, whereas two extinction events take
place around generation 420 and $1\,150$ respectively.

The model can reproduce a number of other features which can be observed
in nature. In particular, let us consider $D>2$ cells on a row.
It is then possible to have a regime in which the two terminal
cells host mutually sterile populations, while they are
each interfecund with the populations in the immediately neighbouring cells.
If we now allow migration between the terminal cells,
we are led to the so-called ``circular invasion''
(see, e.g., Mayr 1963, Grant 1991),
which has been considered as one of the main arguments for
the relevance of geographic speciation. 

Such a situation is shown in fig.~4. We consider a system of
5 islands with $M=100$ individuals per island placed on a row.
Initially individuals can
move only between neighbouring islands, and the first one
does not communicate with the last one. However, migration
between the first and the last islands is allowed starting
from generation 1500.
In the figure the number of potentially
fecund pairs between neighboring islands
is plotted against time.
Immediately after the opening of the
``passage'' between the first and the last island
this number is close to $10\,000$ for all islands.
However, between generation 2500 and 3250
the number is close to zero between island 1 and 2,
while it is close to the maximum everywhere else.
Therefore, during this time, the populations
of islands 2 to 5 are interfecund with their neighboring
ones (and 5 with 1), but most individuals in island
1 are not interfecund with those of island 2.
In this situation, the concept of biological species as a
well-defined set breaks down. However, this situation
is not stable. In the run shown, this genetic fracture is reabsorbed
around generation 3500, whereas two new barriers
(between islands 3 and 4 and 4 and 5) nucleate immediately
afterwards. It would be interesting to analyse the dynamics
of these fracture.

The behaviour of the system in this regime cannot be predicted
with certainty in a wide region of parameters. 
The same system can exhibit different regimes
in different runs of the process.
The reproductive
fracture between neighbouring populations
can be healed, or a new fracture can be spontaneously
arise---leading to the coexistence of well-defined species---or
the ``circular invasion'' situation can remain indefinitely.

We have discussed a simple model of species formation which
takes into account the geographic distribution of the
populations. We have shown that, in spite of its simplicity,
it may reproduce some intriguing features of the species distribution
observed in nature, and that it can justify the relevance
of geographic isolation as one of the main triggers
of species formation.
\ack
L P thanks B Derrida and P C Higgs for illuminating discussions.

\references
\refjl{B Derrida and P C Higgs 1991}{\JPA}{24}{L985--L991.}
\refjl{B Derrida and P C Higgs 1992}{}{}{}
\refjl{B Derrida and L Peliti 1991}{Bull. Math. Biol.}{53}{465--523.}
\refbk{V Grant 1991}{The evolutionary process}{(New York: Columbia)
p 240--241.}
\refbk{E Mayr 1963}{Animal species and evolution}{(Cambridge, Mass.:
Harvard) p 507--512.}
\refjl{M Serva and L Peliti}{\JPA}{24}{L705--L709.}

\figures
\figcaption{Genetic overlaps within one island (squares) and
between the two islands (circles) plotted agains the mutation rate $\mu$
for a population of 200individuals, with a migration rate $\epsilon=0.05$.
The threshold $q_0$ is set at 0.2.
Averages over $10\,000$ generations. The continuous and broken
line represent respectively the predictions for $Q$ and $P$
in the absence of the threshold (eqs.~(6,7)).}
\figcaption{Overlap variances $\delta$ (circles) and $\Delta$
(squares), as defined by eqs.~(9,10), for the population of
fig.~1. The arrow indicates the point in which
$P$ falls below the threshold.}
\figcaption{Number $\nu$ of potentially fecund partners per individual
in a population of 400 individuals distributed in two islands.
Most of the time one observes two values, whose sum is equal to 400,
showing that the population splits in two coexisting well-defined
species. The quantity becomes more widly distributed in
correspondence with speciation events, e.g., near $t=300$ or
$t=700$, which are both followed by the coexistence of {\it three\/}
species. Extinctions occur near $t=420$ and $t=1150$.}
\figcaption{Number of potentially fecund pairs in neighboring islands
versus time for a system of 5 islands, each containing 100
individuals. Initially emigration is inhibited between
islands 1 and 5 (and viceversa). Emigration is then
allowed, starting from generation 1500. Emigration
probability: $\epsilon=0.05$, mutation rate $\mu=0.025$,
fecundity threshold $q_0=0.33$.}
\bye